\documentclass[aps,prb,twocolumn]{revtex4}
\usepackage{graphicx}

\begin{document}
\title
{Physics of the RVB (Pseudogap) State in the Doped Mott 
Insulator:  Spin-Charge Locking
}
\author{P. W. Anderson }
\affiliation{Department of Physics, Princeton University, Princeton NJ 08544}
%\pacs{}
\date{\today}
\begin{abstract}
The properties of the pseudogap phase above $T_c$ of the high-$T_c$ cuprate superconductors are described by showing that the Anderson-Nambu $SU(2)$ spinors of an RVB spin gap ``lock'' to those of the electron charge system because of the resulting improvement of kinetic energy. This enormously extends the range of the vortex liquid state in these materials.  As a result it is not clear that the spinons are ever truly deconfined. A heuristic description of the electrodynamics of this pseudogap-vortex liquid state is proposed.
\end{abstract}
\maketitle
The hole-doped cuprate superconductors have, over the past  decade, been shown to exhibit an unusual phase in the region of their phase diagram immediately above the superconducting ``dome'' in temperature, and especially for underdoped compositions.  This region has been dubbed variously the ``pseudogap'', ``spin gap'', and (by those unafraid of committing themselves) the ``RVB'' or resonating valence bond region.  Fig. 1 shows a generalized phase diagram summarizing the fairly generally accepted phase boundaries between regions of the complex behaviors exhibited by these fascinating systems, and Fig. 2 (due to Ong and Wang \cite{xu_00}) is an experimental delineation of the regions in which this phase exhibits most clearly the behaviors I will discuss later, namely a pronounced  vortex Nernst effect and nonlinear diamagnetic susceptibility, and momentum-dependent energy gap around the Fermi level, in addition to the pseudogap in density of states and magnetic gap known to exist over the wider region below the ``pseudogap temperature'' $T^\ast$. 

In a sense, the existence of a phase of some such sort was predicted quite long before it was observed, in that it was postulated a few months after the discovery of the high Tc cuprates that a quantum liquid of singlet pairs of electrons (``valence bonds'') might exist in an $S=1/2$ Mott insulating system, from which the high $T_c$ superconductor would develop \textit{via} doping \cite{pwa_87}.
Baskaran \textit{et al} \cite{bza},
and Fukuyama \cite{suzumura_88}
even drew schematic phase diagrams for such a system which remotely resembled Fig 1.
In a remarkable and prescient paper \cite{kotliar_88}, Kotliar and Liu, using slave boson methods, actually predicted the existence of a pseudogap phase very much resembling that derived by simpler methods in the present work . But of course, 
they could not compare their results with experiment.

The existence of the pseudogap was first seen in infrared data 
but not recognized \cite{thomas_88}, 
and demonstrated by magnetic \cite{alloul_89}, 
tunneling \cite{renner_98}, thermal \cite{loram_94}, 
ARPES \cite{marshall_96} and other experiments over the years.  
Here I particularly want to bring out two observations: first the data of Ong \textit{et al} on the giant Nernst effect \cite{xu_00}
and the nonlinear diamagnetic susceptibility \cite{ong}
which demonstrate that over a considerable region this phase exhibits a kind of supercurrent in a magnetic field;  and second the data of Timusk \cite{puchkov_96}
which show a marked lengthening of the conductivity relaxation time.

The theory I am going to present for this state starts from the RMFT (renormalized mean field theory) of the superconducting  ground state and uses this as an effective Ginzburg-Landau energy controlling the fluctuations of the variational and other parameters of the state. The variational ``gap equations'' which determine  the energies of the excitations and the energy of the state are \cite{zhang_88, atoz}
\begin{eqnarray}
\Delta_k &= & g_J J \sum_{k'} \gamma_{k-k'} \frac {\Delta_{k'}}{2E_{k'}}\nonumber \\
E^2 _k &= & \xi^2_k + \Delta_k^2 \nonumber \\
\xi_k = g \epsilon_k + \varsigma_k &= & g \epsilon_k +
g_J J \sum_{k'} \gamma_{k-k'} \frac {\xi_{k'}}{2E_{k'}} \nonumber \\
\label{gapeqn}
\end{eqnarray}

Here, $E_k$ is the quasiparticle energy, 
its band-theoretical kinetic energy referred to the Fermi energy, 
$g$ is the kinetic energy renormalization factor due to the modified occupancies
($g = 2x/(1+x)$, $x$ the doping), and $g_J$ the renormalization factor for $J$. 

The physics of strongly-correlated electrons is plagued by the tendency of limiting cases to exhibit special behaviors.  The present case is confused by the fact that the antiferromagnetic insulator is marginally lower in energy than the RVB one, at exactly zero for $g$.  Nonetheless we choose to ignore that fact and treat the problem using the equations (\ref{gapeqn}) in the approximation $g \approx 0$, since the kinetic energy gained in the superconducting-RVB state for $g$ finite soon outweighs this small difference.

The lowest energy translationally invariant state is given by 
these equations with $\varsigma_k \propto \cos k_x + \cos k_y$ and 
$\Delta_k \propto \cos k_x - \cos k_y$ 
or \textit{vice versa}.  As was emphasized in the early work \cite{pwa_87}, \cite{zhang_88},  the solutions, and even the equations, are not unique. It is most evident that  $\varsigma$ and  $\Delta$ may be interchanged, or in fact any orthogonal transformation on the two is allowed. (It is, however, only in the nearest neighbor square lattice without doping that they need have the same magnitude.)
But this is only a symptom of the gigantic symmetry of the solutions, which follows from the fact that the Fermion representation on which they are based is enormously overcomplete.  This overcompleteness generates a local  $SU(2)$ symmetry \cite{pwa_87},  which follows from the fact that a spin up hole and a spin down electron create the same state of the insulating Heisenberg model in Gutzwiller projection.  This symmetry is a true gauge symmetry, in that the states generated by it are physically the same state.  For translationally invariant, singlet  BCS states like (\ref{gapeqn}), the gauge symmetry implies symmetry under $SU(2)$ rotations operating on the three Anderson-Nambu pseudospin vectors $\tau_1$, $\tau_2$, $\tau_3$, 
multiplying respectively the real and imaginary anomalous self-energies and the kinetic energies of a gas of ``spinons''.  

But as I said, this is only a tiny fraction of the available representations.  Also equivalent are states describable as a staggered flux phase, a true $\pi$-flux phase, a $d$-density wave, \textit{etc.}, if one allows translationally non-invariant solutions 
or $T$ non-invariant ones.  What was striking about all good RVB solutions was that they exhibited four nodes at  ($\pi /2,\pi/2$) in the spinon energy as a function of momentum, and this is simply because of their equivalence.  The physics behind this fact is that in order to describe a projected state each self-energy must average to zero hence have a line node; and that the optimum BCS solution must have as little variation of $|\Delta|^2$   
as possible, hence $\Delta$ should be the sum of orthogonal functions.  Without being able to supply a mathematician-style proof, this seems to tell us that RVB's will always have nodes in their spectra, even though there is no Goldstone-type argument for nodes in terms of broken symmetry.

Now we allow g to become finite.  Our approach is to set up a second triad of
$\tau$ 's, thinking if you like of $g$ as defining the $\tau_3^\prime$ direction in this second space.  In the RMFT ground state wave-function written as \cite{pwa_condmat},
\begin{equation}
\Psi_G = \prod_k (gu_k + v_k c^\dagger_{k\uparrow}c^\dagger_{-k\downarrow} |0\rangle~,
\label{wavefn}
\end{equation}
the phase $\Phi$ of the pair wave-function is normally taken to be the phase of $v/u$.  Our procedure will be to ascribe the phase to $g$ and leave $u$ and $v$ real. We can think of $g$ as a vector in the $\tau^\prime$ space which rotates around in the $\tau_1^\prime ,
\tau_2^\prime$ 
plane, thereby defining the third direction. Its meaning at least in the wave function (\ref{wavefn}) seems to be that it is the hole pair amplitude, which in the ground state is Bose condensed.

We now consider the two triplets of vectors $\tau$ and $\tau^\prime$   to be oriented at arbitrary Eulerian angles with respect to each other in the abstract space.  The selfconsistent solutions for $\varsigma$ and $\Delta$ depend only weakly on these angles for small $g$, so a good approximation to the effective Hamiltonian for quasiparticles is then 
\begin{eqnarray}
H &=& \sum_k \left[g\epsilon_k + (\tau_3^\prime \cdot \tau_2) \varsigma_k +
(\tau_3^\prime \cdot \tau_1) \Delta_k \right](n_k+n_{-k}-1) \nonumber \\
& & \mbox{} + \left[
(\tau_1 \cdot \tau_1^\prime)\Delta_k + (\tau_1^\prime \cdot \tau_2) \varsigma_k
\right]c^\dagger_{k\uparrow}c^\dagger_{-k\downarrow} + {\rm h.c.} \nonumber \\
\label{hamiltonian}
\end{eqnarray}                    
Here we are using a kind of Eilenberger method  where we derive an effective free energy as a functional of the parameters in the one-electron Hamiltonian which leads to the gap equation-the most important of these parameters being the angles between the two triads of vectors.  The ``gaps'' themselves, $\varsigma$  and $\Delta$ , do not vary much at small $g$, though $\Delta$, which will become the superconducting gap, will of course decrease appreciably toward optimal doping.  The interaction term in the Eilenberger energy, which will be of the form $(\varsigma^2 + \Delta^2)/J$ , 
will clearly not change much as a function of angles, so that mostly we have to consider the effective one-quasiparticle energy. The ``extended $s$-wave'' solution
$\varsigma_k$ to the RVB gap equations strongly resembles the one-particle kinetic energy-which is not terribly surprising since $J$ is a second-order consequence of the kinetic energy, and therefore the interaction term naturally factorizes  into terms resembling the kinetic energy.  (What may be more germane to the special nature of the cuprates is the existence of the second, orthogonal, $d$-wave solution.)  Therefore it becomes quite obvious that when $\tau_2 \parallel \tau_3^\prime$,
the kinetic energy is very considerably enhanced,  because the coefficient $\xi_k$, 
which is the multiplier of the first term in (\ref{hamiltonian}), then becomes the sum of two terms of the same structure and the same sign.  (This was of course the choice which was made in the original paper on the RMFT \cite{zhang_88}.)  Therefore with this choice we maximize the quasiparticle energies, which necessarily minimizes the total energy. The ``locking'' is illustrated in Fig. 3.

This is the central result of the theory: that upon doping the  $\tau_2$
axis of the RVB triad locks to the $\tau_3^\prime$
axis of the Nambu triad of the charge degrees of freedom of the real electrons,  so that the  $\tau_1$ axis necessarily lies in the $\tau_1^\prime$, $\tau_2^\prime$
plane and $\Delta$ serves as a ( necessarily \textit{real}) anomalous self-energy for the actual electrons, which live on the unprimed triad.  The reason for the name 
``spin-charge locking'' is the similarity to the mechanism of 
``color-flavor locking'' of Wilczek \cite{alford_99},
which in turn is based on the 
``spin-orbit'' locking mechanisms of Leggett for He-3 \cite{leggett_73}.   
But the analogy cannot be carried too far-the peculiar gauge symmetry of the RVB is unique.

A second scale is the energy which maintains the broken true phase symmetry.  The phase in the $\tau_1^\prime$, $\tau_2^\prime$
plane is meaningful only as a relative variable and acquires no stiffness from $J$-
a fact which is equally true for conventional superconductors, for which the stiffness
$\rho_s (\nabla \phi)^2$  
is independent of gap parameters or interactions. The loss of this stiffness may formally be seen as an ``unlocking'' process also, but is not in principle different from a conventional $T_c$. There are therefore two transitions and three different phases:  the low temperature phase is the true superconductor where the two triads may be thought of as locked together; this undergoes loss of phase coherence at Tc, but the $\tau_3^\prime$
and $\tau_2$   
directions remain locked together in the spin-charge locked pseudogap phase, so that the fluctuating anomalous self-energy  remains of d symmetry and has nodes along 
the $(\pi,\pi)$ line; and finally, complete unlocking where there may be a pseudogap but it does not show momentum dependence-there is no connection between the spin gap structure and the hole or particle nature of the excitations.  It would be only in this unlocked phase that we should think of deconfined spinons.

The transition temperature for the ``unlocking'' transition may be estimated from the energy involved and seems unlikely to be sharp.  It must lie between $T_c$ , where the phase coherence dies, and $T^\ast \approx J -xt$,
where the RVB gaps appear \cite{pwa_02}, and we conjecture that it is the 
crossover  or ``onset'' identified by Ong \textit{et al}. 
The energy involved is relatively easy to estimate for low doping, 
where $gt \ll \varsigma, \Delta$.
Here the average kinetic energy of a state with arbitrary orientation of the
$\tau$'s is zero, since there will be no correlation between kinetic energy and occupation; while with the locked configuration , the kinetic energy per electron will be of order
$gt$.  We conjecture that the locking temperature $T^{\ast\prime}$ will be of order $gt$, so that it will rise more steeply than $T_c \approx g \Delta$ . This agrees roughly with Ong's observations (see Fig. 2).  This rise will stop before 
$gt \approx \Delta \approx (J-gt)$, where the two contributions to $\xi$  become comparable; this will come at about $1/2$ of the maximum doping.  From this point on the calculation of the energy becomes quite complicated and we will postpone it to a later paper. Kotliar and Liu suggest that the onset may not change much with further doping \cite{kotliar_88}.

The electrodynamics of the partially locked state is in principle the same as that of the vortex liquid state, in terms of symmetry.  It has in common with that state that there is always an anomalous amplitude, \textit{i.e.}, a ``gap'', but that the phase of this gap is fluctuating freely; there is a classical liquid of vortex lines.  Not much thought has been given to the responses of such a liquid, particularly not to the diamagnetism, which must approximate that of the Abrikosov state at the melting line;  how it diminishes from this has not been discussed. 

I have concluded that an approximation to its behavior may be obtained in the following way.  The instantaneous response to an electromagnetic field is identical to that of an ordinary superconductor, but the current-current correlation decays with a finite correlation time $\tau$.  From this picture we can predict two responses. The superconducting response to an electric field is the acceleration equation: 
$\frac {dJ}{dt} = \rho_s E$
; correspondingly, for the locked but non-superconducting state we should have 
\begin{equation}
J_s = \rho_s \tau E;~~\sigma_s = \rho_s \tau~.
\label{diamag}
\end{equation}
We have no apriori reason to select a magnitude for 
$1/\tau$, but it certainly should remain less than $\Delta$, otherwise the picture is meaningless;  and experimental data on infrared response in the RVB region suggest that the vortices freeze at $1/\tau = T = T_c$ \cite{homes_04}.

The diamagnetic response is more interesting.  Our picture of the pseudogap-vortex liquid state is that the order parameter remains non-vanishing but has no long-range phase order. If one were to introduce a flux quantum instantaneously at a point one would induce supercurrents throughout the sample, 
including the long-range current $(\hbar /m) \rho_s \nabla \phi \propto 1/r$.
The divergent energy due to such currents means that in a superconductor they must be screened out by a surface current, which is responsible for the magnetization of the superconductor.  It is these magnetization currents which correspond to the accelerated supercurrent in an electric field and exhibit a delta-function response
$\delta(\omega) \to 1/(i\omega + 1/\tau)$
when the gap fluctuates thermally with a correlation time $\tau$. This response is proportional to the magnetization which for low fields varies as
$B \ln \frac{H_{c2}}{B}$. Thus the susceptibility
\begin{equation}
\chi \propto \tau \ln \frac{H_{c2}}{B}~,
\label{susc}
\end{equation}
will have a logarithmic divergence as $B \to 0$. 
This singularity was first observed by Ong and Lu \cite{ong}, 
in fact it was the search for a mechanism for this behavior which led to the present line of thinking (although the observed singularity seems to be more precisely described by a power law.)  They also observe that the upper critical field $H_{c2}$
is continuous through the melting line and remains large and finite to temperatures where the susceptibility can no longer be measured.  This field must be that where the charge becomes unlocked from the RVB, since I can't see how the field would have much effect on the RVB.  It is hard to find any other explanation for the remarkable fact that the susceptibility remains singular over such a wide range of temperatures;  no explanation in terms of critical behavior seems viable.

It is possible to improve somewhat on the simple expression (\ref{susc}), which is meant only to show the nature of the singularity at $B=0$. In the $d$-wave superconductor, there is a distribution of gaps and therefore we may approximate its response by summing over a distribution of $H_{c2}$'s. To get a notion of the form of the variation, let us assume that the distribution of the gaps is uniform; if we do this, we find that the expression for the magnetization becomes,
\begin{equation}
M \propto B~(\sqrt B -1-\ln \sqrt B)~,
\label{magnetizn}
\end{equation}
where $B = H/H_{c2{\rm max}}$. This is illustrated in Fig. 4 and its resemblance to some of Ong's data is evident.

In summary, I have proposed a theory which accounts for most of the anomalous experimental facts about the pseudogap state, and connects rather seamlessly to the only successful microscopic theory of the superconducting state.  Neither theory places much emphasis on the complications of the various inhomogeneous phases which tend to occur in these systems but which seem to involve smaller energies and weaker perturbations than the more striking and universal effects we discuss.  (In fact, at least one such phenomenon, the ``checkerboard'', seems to receive a natural explanation within this theory \cite{pwa_condmat}.) 
The crucial step seems to be the idea of visualizing the RVB as a separate entity in the spin Hilbert space which is locked to the electron charge Hilbert space by a relatively weak force.  This makes irrelevant the many ``ghost vacua'' of the spin system which have led other workers astray.  Yet when the spin system chooses a physically different ``vacuum''-the antiferromagnetic state, for instance-the whole complex of superconductivity completely disappears \cite{bozovic_03}. On the other hand, as Ogata has shown, it is possible for antiferromagnetic order to coexist with $d$-wave superconductivity \cite{himeda_99}.

In terms of the symmetry classification of phases, on which often great emphasis is placed, I have not discovered any ``new'' entities.
The superconducting state is only unconventional in that its quasiparticles are not hole-particle symmetric.  The pseudogap phase is ``only'' an enormous extension of the vortex liquid. Yet each exhibits new and unexpected physical phenomena.  One is tempted to speculate that through too much emphasis on mathematical categorization physical reality has been neglected.

%\bigskip
I would like to acknowledge the importance of continued contact with the experimental results and valuable interpretive ideas of N. P. Ong and his group, and also continued discussions with V. N. Muthukumar. G Baskaran, G Kotliar and P A Lee as founders of the gauge theories, as well as  C Gros, F C Zhang, M Randeria, and N Trivedi as originators and resuscitators of the RMFT theory, are not adequately acknowledged by the text references.

\pagebreak
\clearpage

\begin{figure}
\begin{center}\includegraphics[width=0.9\columnwidth]{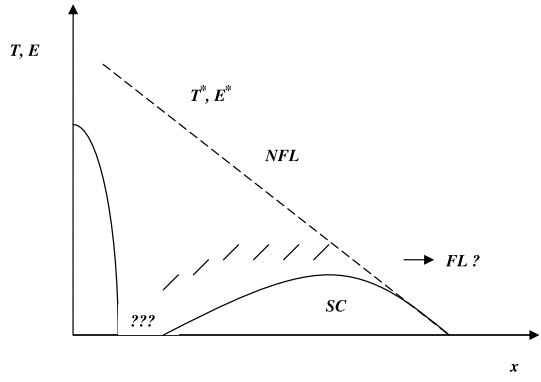}\end{center}
\caption{Generalized phase diagram of the cuprate superconductors.}
\end{figure}

\begin{figure}
\begin{center}\includegraphics[width=0.95\columnwidth]{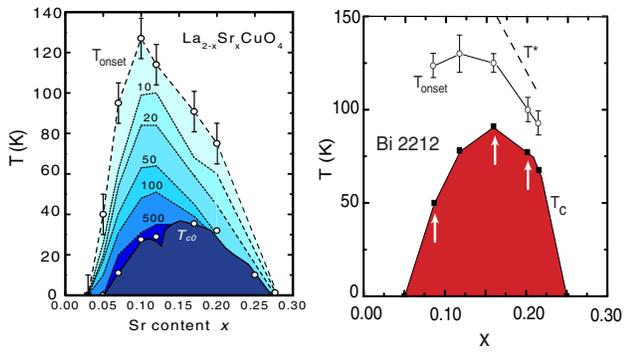}\end{center}
\caption{The ``dome'' and the region of observation of a vortex Nernst effect, as a function of doping for two superconducting systems.}
\end{figure}

\begin{figure}
\begin{center}\includegraphics[width=0.9\columnwidth]{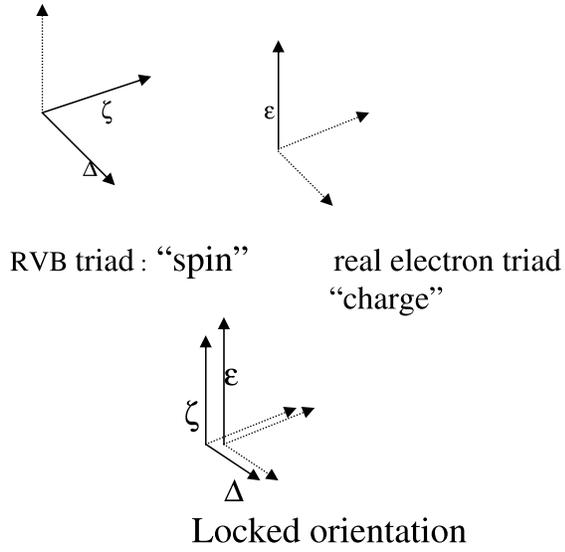}\end{center}
\caption{The charge and spin pseudovector triads, and the relative orientation when ``locked'' (in the superconducting state.)}
\end{figure}

\begin{figure}
\begin{center}\includegraphics[width=0.9\columnwidth]{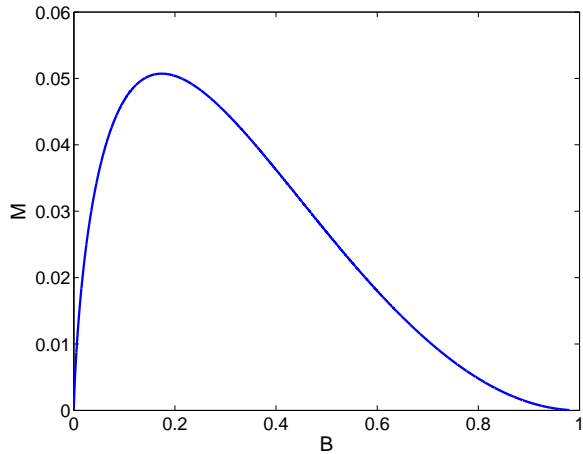}\end{center}
\caption{
Hypothetical variation of diamagnetism with magnetic field in pseudogap phase. Here, 
$B = H/H_{c2{\rm max}}$, and the scale of $M$ decreases as  $\tau(T)$.}
\end{figure}

\end{document}